\documentclass[%
 reprint, amsmath, amssymb, showkeys, aps, pra, letterpaper
]{revtex4-2}
\pdfoutput = 1

\usepackage{graphicx}% Include figure files
\usepackage{dcolumn}% Align table columns on decimal point
\usepackage{bm}% bold math
\usepackage{hyperref}% add hypertext capabilities
\usepackage{float}
\usepackage{subcaption} %  for subfigures environments
\usepackage{comment}
\usepackage{placeins}
\usepackage{lipsum}
\usepackage[export]{adjustbox}

\usepackage{xcolor}

\graphicspath{ {pictures/} }

\begin{document}

\preprint{APS/123-QED}

\title{Image Processing Methods Applied to Motion Tracking of Nanomechanical Buckling on SEM Recordings}

\author{Ege Erdem{$^1$}}
\author{Berke Demiralp{$^1$}}
\author{Hadi S Pisheh{$^1$}}
\author{\\Peyman Firoozy{$^1$}}
\author{Ahmet Hakan Karakurt{$^1$}}
\author{M. Selim Hanay{$^{1,2}$}}%
 \email{Corresponding author. \\
selimhanay@bilkent.edu.tr}
\affiliation{{$^1$}Department of Mechanical Engineering, Bilkent University, 06800, Ankara, Turkey}
\affiliation{{$^2$}National Nanotechnology Research Center (UNAM), Bilkent University, 06800, Ankara, Turkey}

\date{\today}

\begin{abstract}

The scanning electron microscope (SEM) recordings of dynamic nano-electromechanical systems (NEMS) are difficult to analyze due to the noise caused by low frame rate, insufficient resolution and blurriness induced by applied electric potentials. Here, we develop an image processing algorithm enhanced by the physics of the underlying system to track the motion of buckling NEMS structures in the presence of high noise levels. The algorithm is composed of an image filter, two data filters, and a nonlinear regression model, which utilizes the expected form of the physical solution. The method was applied to the recordings of a NEMS beam about 150~nm wide, undergoing intra-and inter-well post-buckling states with a transition rate of approximately 0.5~Hz. The algorithm can track the dynamical motion of the NEMS and capture the dependency of deflection amplitude on the compressive force on the beam. With the help of the proposed algorithm, the transition from inter-well to intra-well motion is clearly resolved for buckling NEMS imaged under SEM.

\begin{description}
\item[Keywords]
NEMS, buckling, nonlinear nanomechanical systems, post-buckling analysis, image processing, SEM
\end{description}
\end{abstract}

%\keywords{Computer vision, Bucking, NEMS}%showkeys at doc.class to show

\maketitle

\section{\label{sec:1}Introduction}

The prevalence of NEMS devices has seen a sharp increase recently for potential applications in sensing (e.g. charge, mass, stiffness) and mechanical computation \cite{A_Dominguez, A_Matheny, A_Gil, A_Erdogan, A_Irek,  nems_hanaygroup, nems_props, Jin2023}. Their high force sensitivity, operation frequency, mechanical quality factor, integrability and low power consumption level played a major role in this development \cite{nems_props, nems_modeling}. NEMS devices also constitute a readily accessible experimental platform for nonlinear dynamics \cite{Samanta2023, B_Samantha, B_Xu, B_Xie, B_Keşkekler, Lifshitz2008}. As a nonlinear phenomenon, buckling \cite{C_Gomez2016, C_Wang2023} has been used to improve some of the aforementioned qualities and also to make mechanical information storage/processing possible \cite{nems_hanaygroup, nemsberke, D_Weick, D_An2018}.

Visual data in nanomechanical buckling is usually obtained using a SEM \cite{nems_imageprocessing}, which results in blurriness for dynamic systems due to low frame rate along with disturbances induced by applied electric potentials \cite{nemsberke}. In some studies, this problem is eliminated by improving the frame rate by reducing the number of scanned pixels \cite{nemsberke, STEM_scanning}, although image denoising methods are also popularly used \cite{merchant2022microscope, atom_imagproc, denoising_STM, SEM_qualityrealtime}. Since object tracking algorithms conventionally used are susceptible to noise, a noise-robust algorithm tuned for tracking the type of curves encountered in nanomechanical motion is required \cite{curve_noise}.

In this paper, we present a set of image processing steps guided by the physical solution of the underlying dynamical system to track the buckling motion of a NEMS beam in a noisy environment. The algorithm starts with denoising filters to obtain a set of filtered points, which are then fed into a nonlinear regression model which mimics the physical post-buckling solution. This way, the post-buckling behavior of the system when confined to a single potential well and during its transition between the two wells can be clearly extracted from the video recordings.

\vspace{-3mm}

\section{\label{sec:2}NEMS Setup}

\begin{figure*}
    \begin{picture}(200,175)
        \put(-155,0){\includegraphics[width=0.39\linewidth]{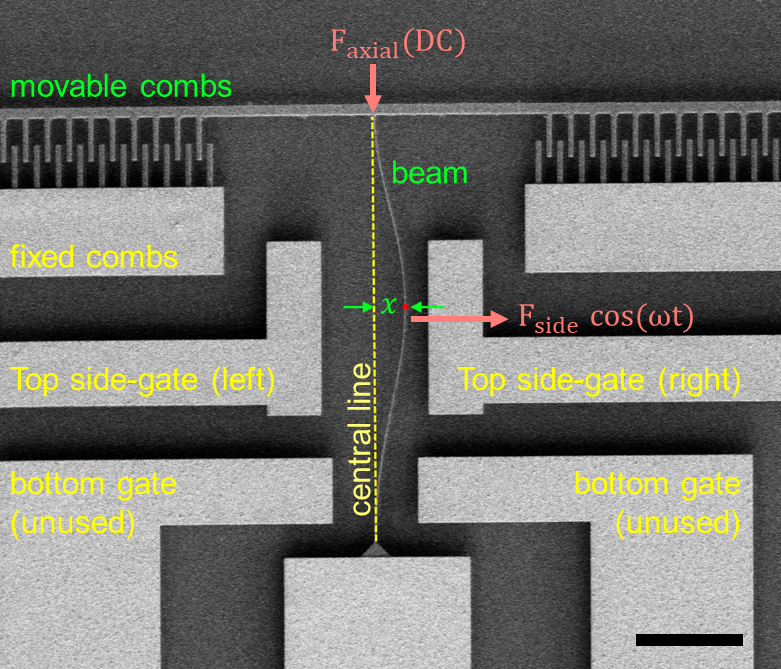}
        \phantomsubcaption\label{sfig:setup}}
        \put(48,82){\includegraphics[width=0.21\linewidth]{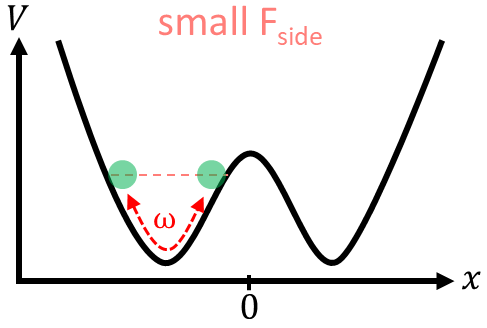}
        \phantomsubcaption\label{sfig:setup_single}}
        \put(48,-9){\includegraphics[width=0.21\linewidth]{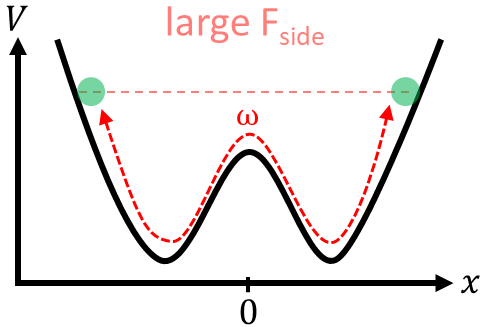}
        \phantomsubcaption\label{sfig:setup_double}}
        \put(170,0){\includegraphics[width=0.35\linewidth]{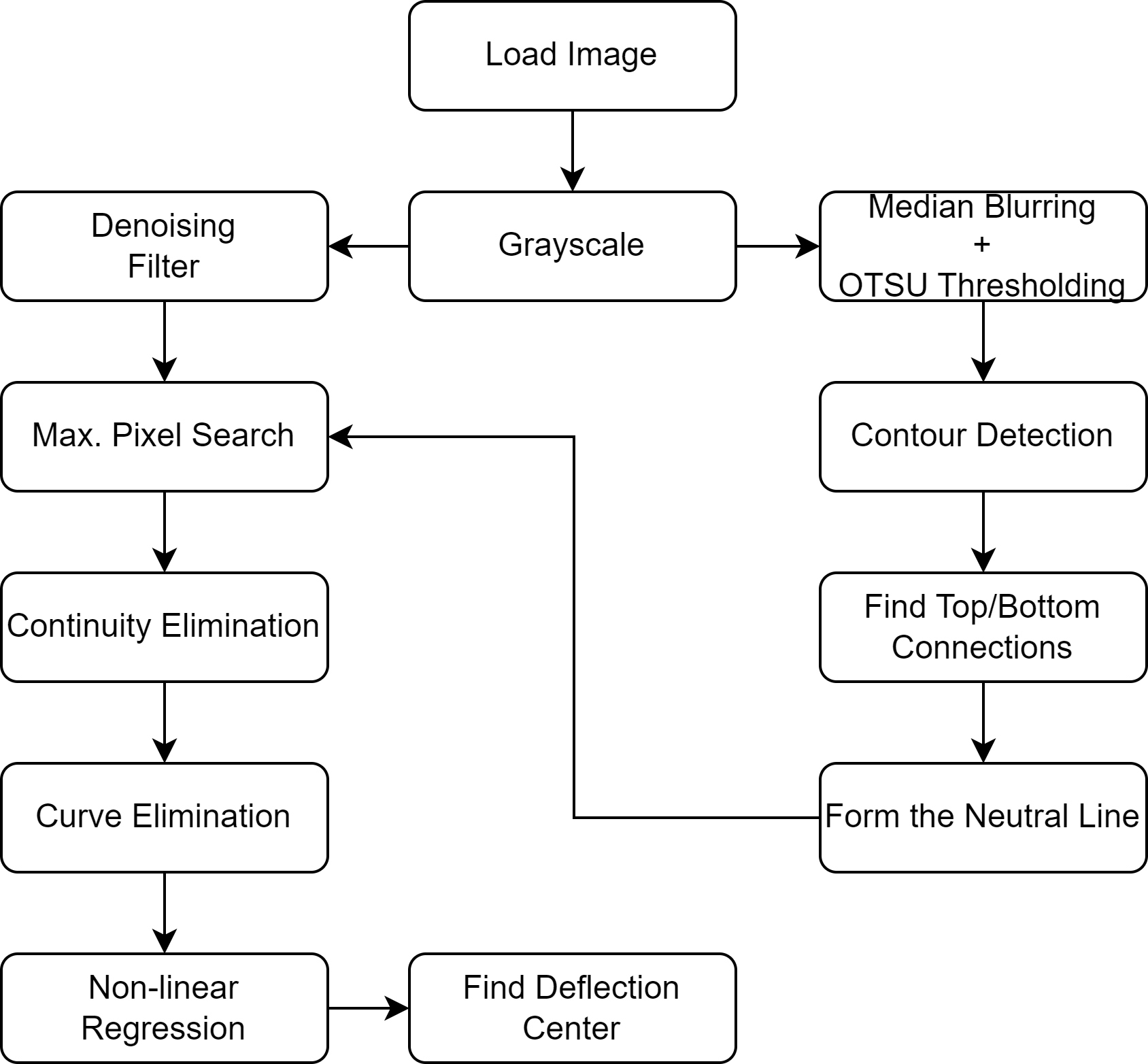}
        \phantomsubcaption\label{sfig:flowchart}}

        \fontfamily{phv}\selectfont
        \put(-153,159){\color{white} \large (a)}
        \put(45,159){\color{black} \large (b)}
        \put(45,70){\color{black} \large (c)}
        \put(170,159){\color{black} \large (d)}
        
    \end{picture}
    \captionsetup{singlelinecheck = false, justification=raggedright}
    \caption{The nanomechanical buckling setup in (a), where the bending magnitude is measured from the central line (scale bar 10 {\textmu}m). (b) Typically, the beam exhibits one-well motion when $F_{side}$ is small, (c) and double-well motion for large $F_{side}$. The diagram describing the algorithm as a whole is shown in (d).}\label{fig:setup_ALL}
\end{figure*}
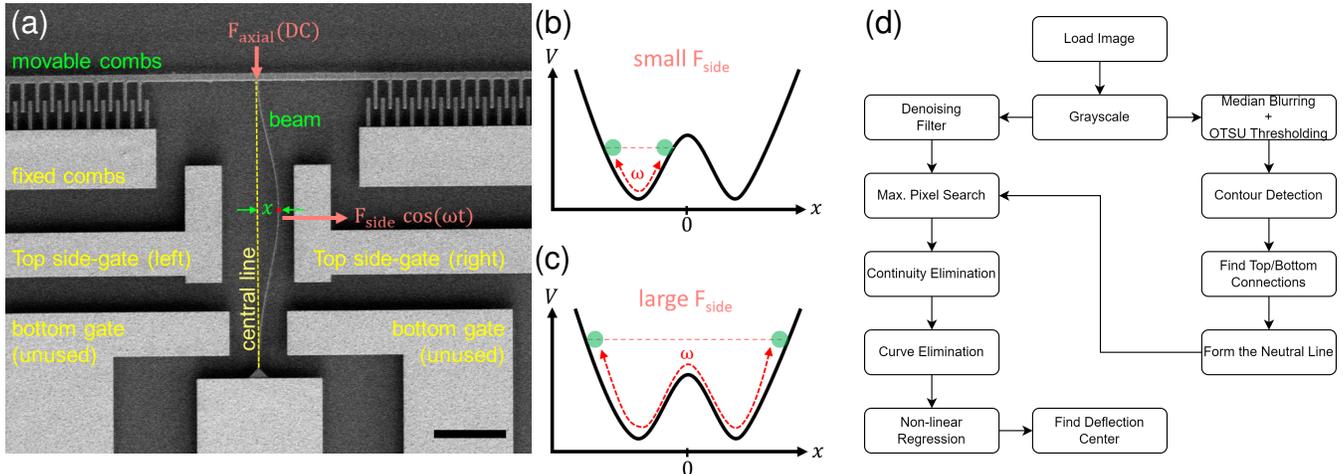

The experimental setup analyzed in this paper is comprised of a nanomechanical beam fixed at the bottom, vertically compressed from the top to achieve critical load, and horizontally actuated from the top side gates (see Fig. \ref{sfig:setup}). The magnitude of the DC voltage across the combs controls the depth of the wells, whereas the AC voltage applied to the top side gate pairs controls the driving force amplitude. At small AC voltages, the system only oscillates within one of the potential wells (see Fig. \ref{sfig:setup_single}); while at larger AC voltages, the system can traverse between the two wells and oscillate back and forth (see Fig. \ref{sfig:setup_double}). The video containing the buckling data is obtained using a SEM~\cite{nemsberke}.

The goal for the image processing algorithm is to track the movement of the beam as the input control voltages (side gate or comb drive) change. The amount of buckling is measured from the point with the maximum orthogonal distance to the imaginary central line that connects the top and bottom connections of the buckling beam.

%\vfill

\section{\label{sec:3}Image Processing Steps}

The image processing for the beam is conducted in two main steps. The first step is to approximately \textbf{locate} the beam within the image; using this information, the second step is to \textbf{track} the oscillation of the beam itself. The overall procedure is summarized in Fig. \ref{sfig:flowchart} and an example output can be observed in Supplementary Video.

\begin{comment}
    \begin{figure}[H]
        \begin{center}
            \includegraphics[width=6.85cm]{pictures/2.1 Flowchart.png}
        \end{center}
        
        \caption{Image processing flowchart. The right branch is mainly related to locating the beam, whereas the left one focuses on tracking the beam.}\label{fig:flowchart}
    \end{figure}
\end{comment}

The location of the beam in SEM recordings does not change much and there is only one beam in the recordings. For these reasons, the algorithm used to locate the beam does not have to be a general-purpose algorithm to find the beam in any given arbitrary image. Rather, its main purpose is to find the connection points of the beam to draw the central line of the unbuckled beam. This way, we can characterize the bending of the beam after tracking all points that belongs to the beam.

\subsection{\label{sec:3-1}Locating the Beam}

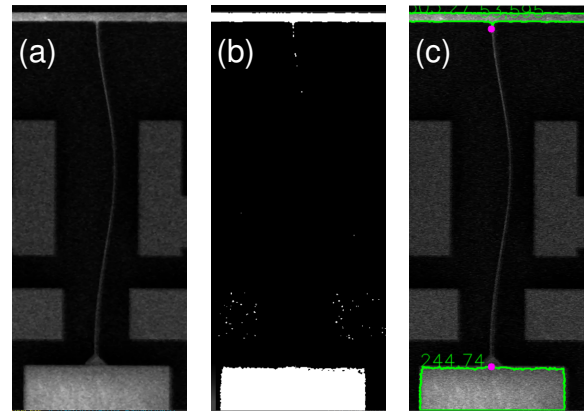
\begin{figure}[H]
    \begin{minipage}{1.3in}
        \begin{picture}(150,150)
            \put(15,0){\includegraphics[width=0.7\linewidth]{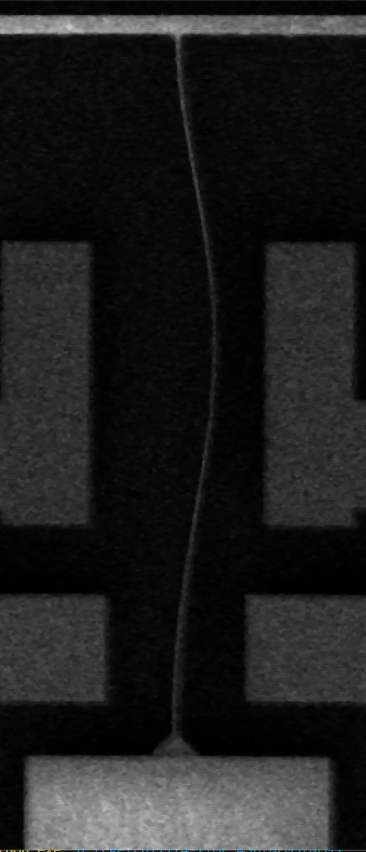}}
            \put(90,0){\includegraphics[width=0.7\linewidth]{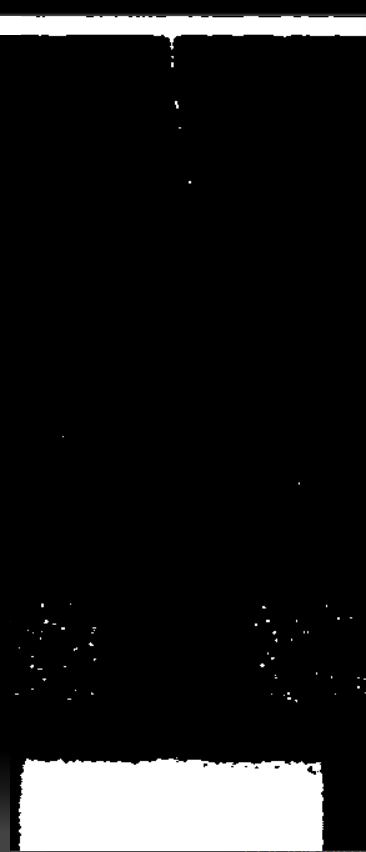}}
            \put(165,0){\includegraphics[width=0.7\linewidth]{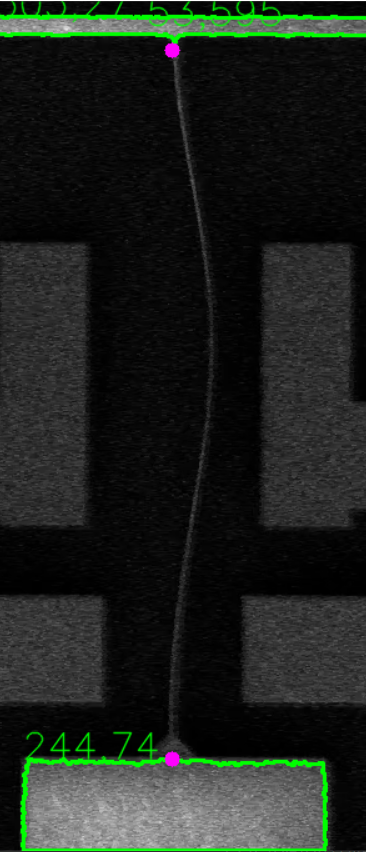}}
            \fontfamily{phv}\selectfont
            \put(17,132){\color{white} \large (a)}
            \put(92,132){\color{white} \large (b)}
            \put(167,132){\color{white} \large (c)}
        \end{picture}
    \end{minipage}
\caption{Connection point finding procedure with contour detection. The input greyscale image (a) is transformed using median blurring and OTSU binarization (b). Then the connected objects are found by contour detection (c). Pink dots on the image denote the center of the clamping region identified by the algorithm.}\label{fig:contour}
\end{figure}

Due to the image shift during the application of AC voltage to the side gates of the buckling device, the clamping regions of the NEMS beam are located first. The background image is separated from the foreground (e.g. control gates, combs) using image binarization via thresholding where the chosen threshold value separates the image between the background and the foreground. OTSU binarization is used for this, which tries every thresholding value between 1 and 255, and picks the thresholding value that minimizes the sum of variance within the background and the variance within the foreground image. In addition, median blurring is applied before OTSU to increase separation accuracy. Finally, the binarized image allows us to use a contour detection algorithm~\cite{contours_suzuki} to find the two contours the beam is attached to (see Fig. \ref{fig:contour}). The center of these contours are aligned with the centers of clamping points of the beam, from which the central line of the unbuckled beam is formed as defined in Sec. \ref{sec:2}.

The two contours here are chosen to track the location of the beam because they have large areas undisturbed by the noise induced during operation. However, applying contour detection to the beam itself is unviable due to the slim profile of the beam and noise.

\vspace{-8mm}

\subsection{\label{sec:3-2}Tracking the Beam}

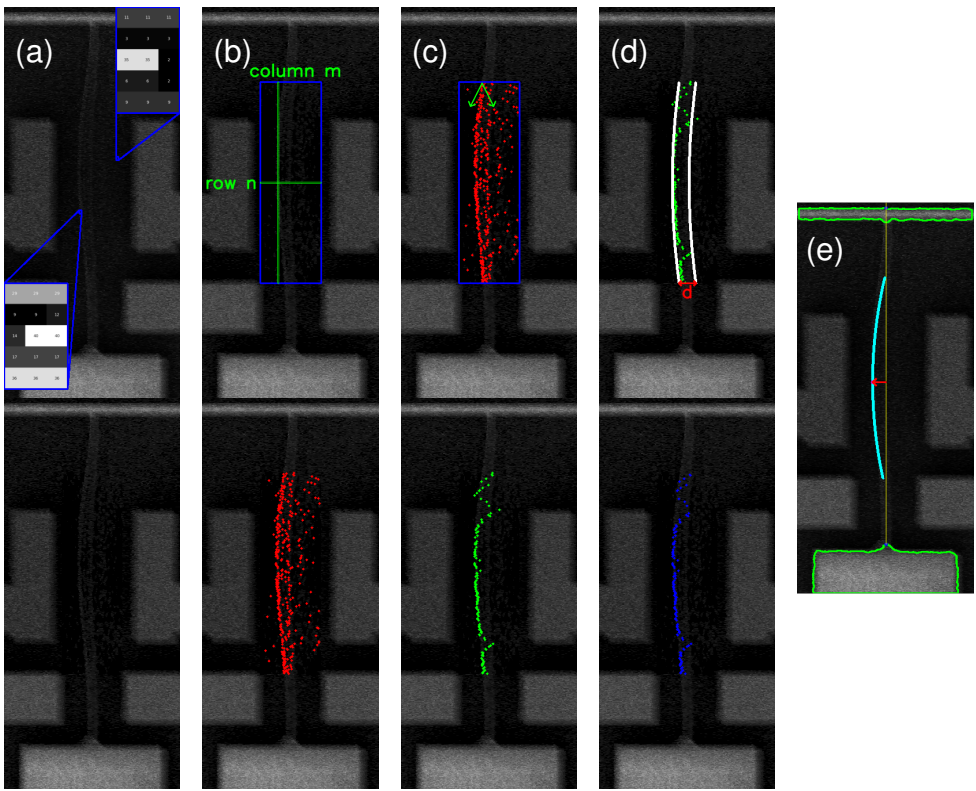
\begin{figure*}
    \begin{picture}(200,300)
        \put(-100,0){\includegraphics[width=0.13\linewidth]{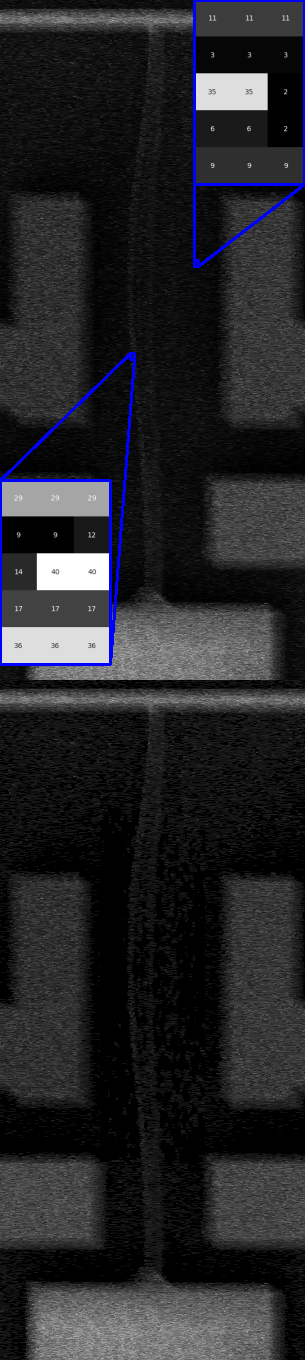}
        \phantomsubcaption\label{sfig:method1}}
        \put(-25,0){\includegraphics[width=0.13\linewidth]{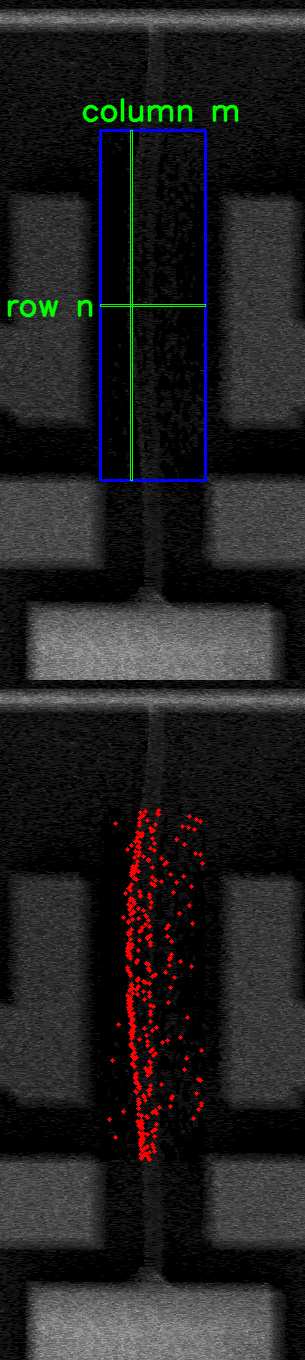}
        \phantomsubcaption\label{sfig:method2}}
        \put(50,0){\includegraphics[width=0.13\linewidth]{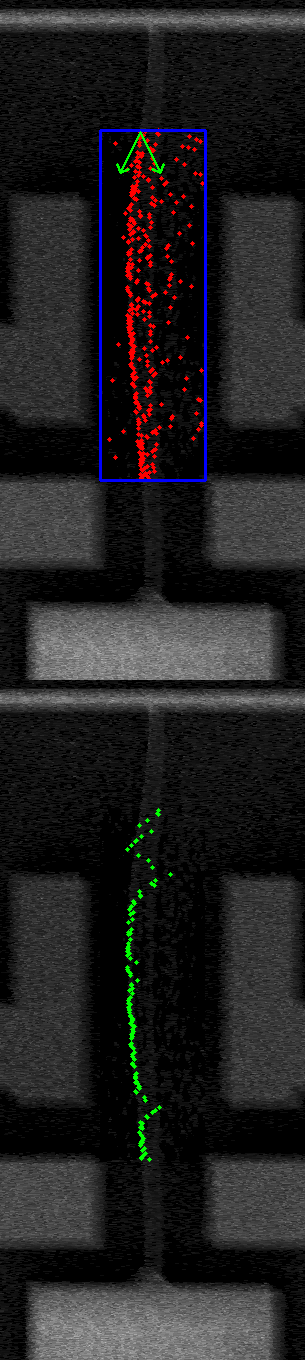}
        \phantomsubcaption\label{sfig:method3}}
        \put(125,0){\includegraphics[width=0.13\linewidth]{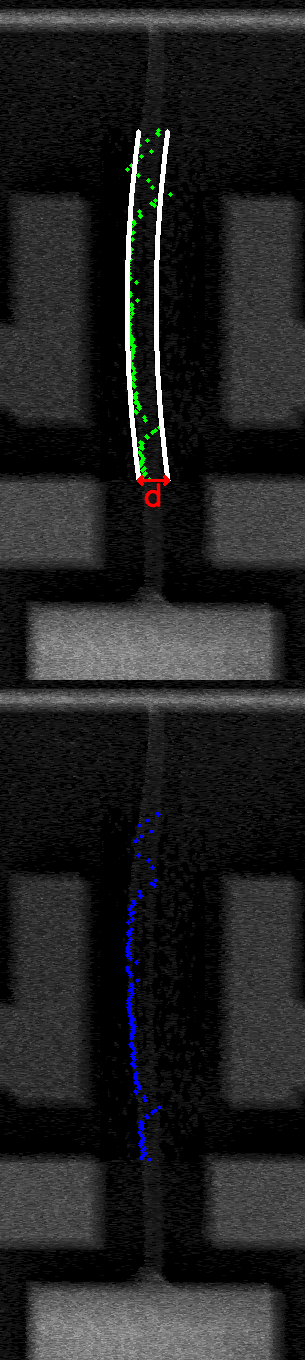}
        \phantomsubcaption\label{sfig:method4}}
        \put(200,0){\includegraphics[width=0.13\linewidth]{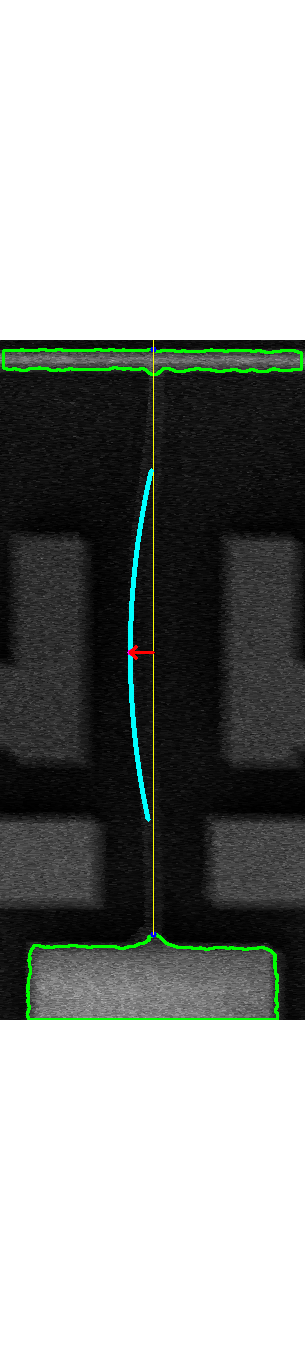}
        \phantomsubcaption\label{sfig:method5}}
        \fontfamily{phv}\selectfont
        \put(-96,275){\color{white} \large (a)}
        \put(-21,275){\color{white} \large (b)}
        \put(54,275){\color{white} \large (c)}
        \put(129,275){\color{white} \large (d)}
        \put(203,200){\color{white} \large (e)}
    \end{picture}
    \captionsetup{singlelinecheck = false, justification=raggedright}
    \caption{Beam tracking procedure as indicated by the method at the top and result at the bottom. (a) Denoising filter applied to the greyscaled image. (b) Row-based maxima search to get initial points. (c) The elimination of erroneous data points by continuity, and (d) by parabolic curve assumption due to the physical solution. (e)~Obtaining the final picture by applying nonlinear regression to the data.}\label{fig:method}
\end{figure*}

After locating the connection points of the beam, the beam is tracked by using a succession of five algorithms, most of which are inspired by the physical solution of the post-buckling beam. A denoising filter is applied first and then positions with the maximum brightness value for each row are taken as data points. These data points are filtered with the assumption that the beam is a continuous parabolic object. Finally, a post-buckling curve is fitted to the remaining data points to select the set of points corresponding to the beam. This procedure is illustrated in Fig. \ref{fig:method}.

%\FloatBarrier
%\clearpage
%\newpage
%\pagebreak
%\pagebreak
%\FloatBarrier
%\newpage

The \textbf{image denoising filter} works with the assumption that the beam itself is physically a continuous object aligned vertically in the recording. Therefore, the pixels connected to the beam will also have nearby bright pixels whereas we expect the noise to exhibit a more random pattern. Applying a convolution filter with the kernel matrix given in Eq.~\eqref{eq:kernel_eq} results in the normalized sum of the nearby pixels for each pixel. Denoising is achieved by applying thresholding to the convoluted output and using the final output as a mask on the original image as shown in Fig. \ref{sfig:method1}. See Appendix \ref{sec:denoise} for a more detailed illustration of the noise in these recordings and the effect of the denoising filter.

\begin{equation}\label{eq:kernel_eq}
    K = \frac{1}{21}{\begin{bmatrix}
        1 & 1 & 1\\
        1 & 1 & 1\\
        1 & 1 & 1\\
        1 & 1 & 1\\
        1 & 1 & 1\\
        1 & 1 & 1\\
        1 & 1 & 1
        \end{bmatrix}}
\end{equation}

The \textbf{row-based maxima search} algorithm works by marking the column with the maximum pixel value for each row as shown in Fig. \ref{sfig:method2}. In perfect conditions, this step alone would be enough; however, the excess noise when AC voltages are applied makes the problem a lot more difficult. Therefore, these data points need to be filtered for a good fit. 

The first filtering step is by the use of \textbf{continuity}. Since the beam is physically a continuous object, the data points for successive rows must be close to one another on the x-axis. For each row, the next rows have a cone of safety. If the succeeding row is not within the cone, it is regarded as a discontinuity and eliminated as shown in Fig. \ref{sfig:method3}. If the succeeding data point is removed for a given row, then the next row will be evaluated with double the margin. This elimination procedure is done from the top and the bottom row simultaneously to ensure that there is no elimination bias in terms of direction. Without this step, there are too many outliers that might mislead the next step (curve elimination) in the presence of noise.

The second filtering step is performed by using a \textbf{parabolic} curve which approximates the post-buckling solution sufficiently enough to represent the shape of the beam. This is achieved by finding two parabolic curves on each side of the beam, having a fixed separation distance $d$, with bending chosen such that the resulting two curves contain the highest number of data points between them. The data points outside of the curves are considered outliers to be removed, as shown in Fig. \ref{sfig:method4}. Setting a high separation distance will not eliminate any points, and the converse will lead to a situation where the majority of the points that are crucial in representing the curve of the beam being eliminated. The parabola as a shape was chosen in this step for easier discretization; however, other shapes can be used as well in similar applications. Without this elimination step, the final curve fitting step usually does not result in a good match and offsets from the beam slightly.

Finally, \textbf{curve fitting} guided by the buckling solution is applied to the remaining data points using nonlinear regression to find parameters that best represent the beam as in Fig. \ref{sfig:method5}  (see Appendix \ref{sec:curvefit}). The fit matches with the beam quite well after filtering the noisy detections.

\section{\label{sec:4}Results}

The image processing algorithm was tested on $1024\times942$ pixel SEM recordings at 10~$s^{-1}$ frame rate. However, the effective resolution of the recording can be stated as $105\times350$ pixel for performance reasons, since the main body of the algorithm was used on only a patch of the recording that contains the beam. Using the scale bar in the microscope image, each pixel corresponds to about 71.4~nm  whereas the beam is a few hundred nanometers wide. Under these conditions, the algorithm manages to perform at about a frame rate of 7~$s^{-1}$ on an Intel i7 processor and the tracking matches the beam perfectly if the noise is not too excessive. Otherwise, there is a slight margin of error of 1-2 pixels.

\begin{figure}[H]
    \begin{center}
        \includegraphics[width=8.2cm]{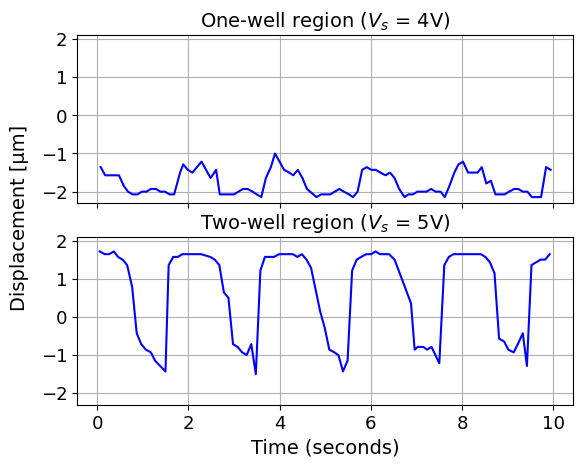}
    \end{center}
    \vspace{-4mm}
    \caption{Behaviour of the nanomechanical beam with varying side gate voltage amplitude and constant comb drive voltage ($V_{pp} = 36V$). For small side gate voltage, the system oscillates within a single well (top panel), whereas increasing the voltage causes the system to travel over the potential barrier.}\label{fig:waves_vs}
\end{figure}

The results of applying the algorithm to recordings with one-well and double-well cases are as shown in Fig.~\ref{fig:waves_vs}. Usually in the recordings, the side gate voltage is the decisive factor for determining the type of dynamical solution (see \cite{nems_hanaygroup} for more details on the behaviour).

\begin{figure}[H]
  %\hspace{0.1cm}
  \adjustbox{minipage=1.3em}{\subcaption{}\label{sfig:testa}}%
  \begin{subfigure}[t]{\dimexpr.99\linewidth-1.3em\relax}
  \centering
  \includegraphics[width=.99\linewidth,valign=t]{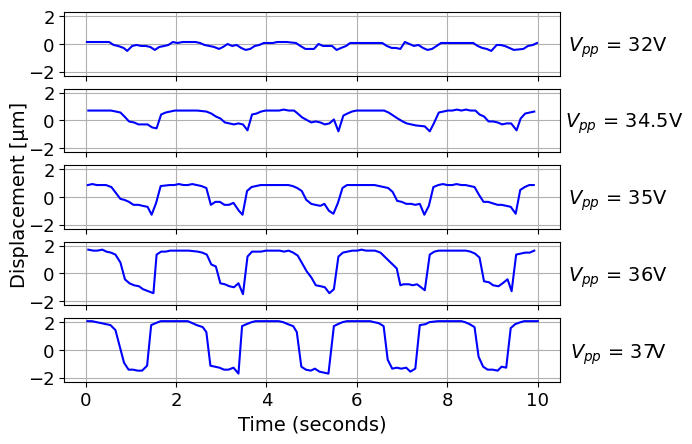}
  \end{subfigure}%
  
  %\hspace{0.1cm}
  \adjustbox{minipage=1.3em}{\subcaption{}\label{sfig:testb}}%
  \begin{subfigure}[b]{\dimexpr.94\linewidth-1.3em\relax}
  \centering
  \includegraphics[width=.98\linewidth,valign=t]{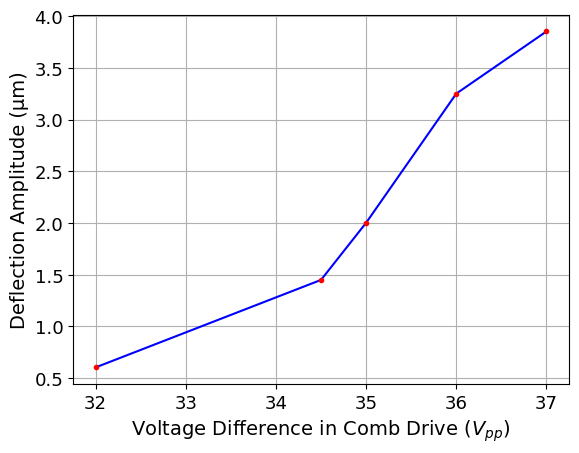}
  \end{subfigure}
  \caption{Behaviour of the nanomechanical beam with varying comb drive voltage and constant side gate voltage ($V_s = 5V$). (a) The beam is in the double well region and the increase in comb drive voltage increases its deflection amplitude as also illustrated in (b).}
  \label{fig:waves_vpp}
\end{figure}

The analysis of the SEM recordings shows that, under a fixed AC amplitude at side gate for inter-well oscillations, the maximum deflection of the beam is approximately proportional to the comb drive voltage as shown in Fig.~\ref{fig:waves_vpp}. In this case, the upper graph represents the situation slightly beyond the buckling. Owing to the small value of compression voltage ($32V$).

\section{\label{sec:5}Conclusion and Future Work}

Application of the proposed image processing algorithm on the SEM recordings clearly shows the relationship between comb drive voltage (offsetting from the buckling threshold value) and beam deflection amplitude. Furthermore, the characteristic motion of the beam is resolved as the side gate voltage changes. However, the algorithm is of limited applicability for recordings with low frame rate or resolution for the beam (which results in more noise). For such recordings, image denoising methods such as wavelet denoising or blind deconvolution might be further explored. Another way to improve image quality would be to obtain higher frame rate recordings by decreasing the SEM scanning area~\cite{STEM_scanning}. The algorithm gets a frame rate of around 7~$s^{-1}$, which is adequate for a non-real-time processing method. 

Compared to using the line mode of SEM (1D scan)~\cite{nemsberke}, this algorithm uses the entire 2D image and marks the shape of the beam as a whole. Therefore, it can be used in analyzing the behaviour of the NEMS beam in more depth as it contains more detailed information about the beam compared to single line scanning. This would be particularly helpful in recordings that demonstrate nonlinear behaviour.

This image processing approach was inspired by the physical description of a continuous beam under buckling and implements its governing equation; however, the overall method can be applied to other physical systems as well.

\begin{acknowledgments}
This work was supported by the Scientific and Technological Research Council of Turkey (TÜBİTAK), Grant No. EEEAG-115E833.

E.E. developed the main image processing framework of the study, analyzed the data and wrote the manuscript; B.D. set up the experimental platform; H.S.P. fabricated the devices; P.F. and A.H.K. obtained the video recording data; M.S.H. supervised the project.
\end{acknowledgments}

\appendix

\renewcommand{\thefigure}{A\arabic{figure}}
\setcounter{figure}{0}
\section{\label{sec:denoise}Noise During Operation and Denoising Filter}

\begin{figure}[H]
    \begin{center}
        \includegraphics[width=7.5cm]{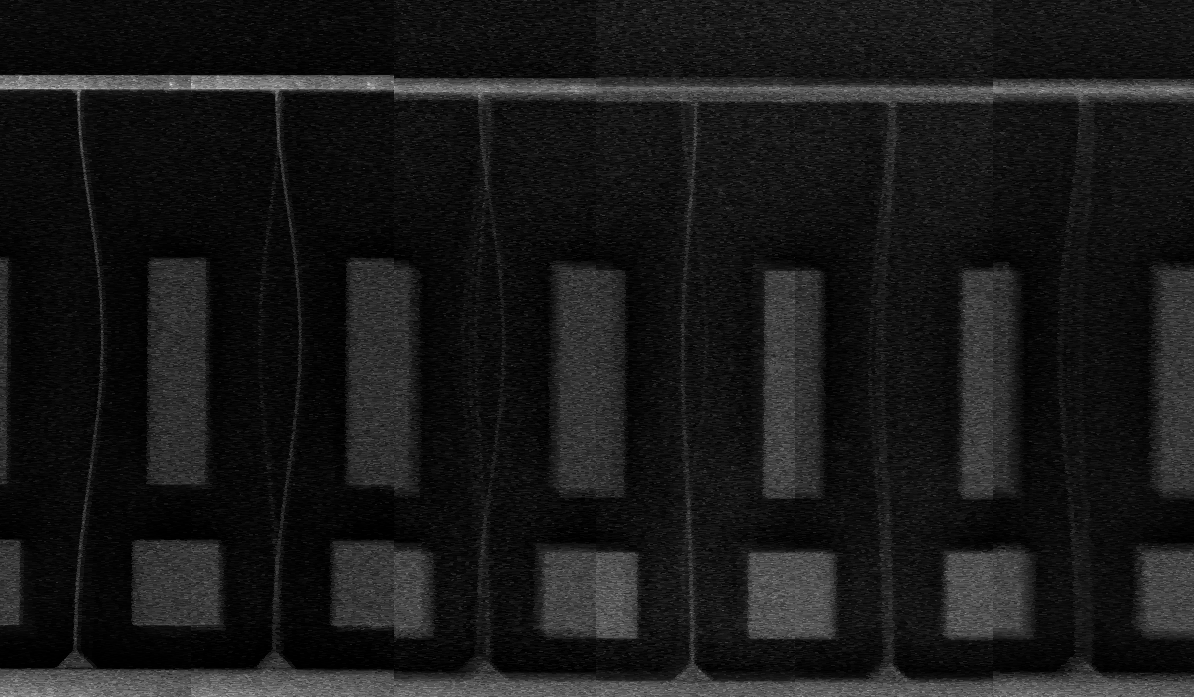}
    \end{center}
    \caption{Noise during operation.}\label{fig:shadowing}
\end{figure}

\begin{figure}[H]
    \begin{center}
        \includegraphics[width=7cm]{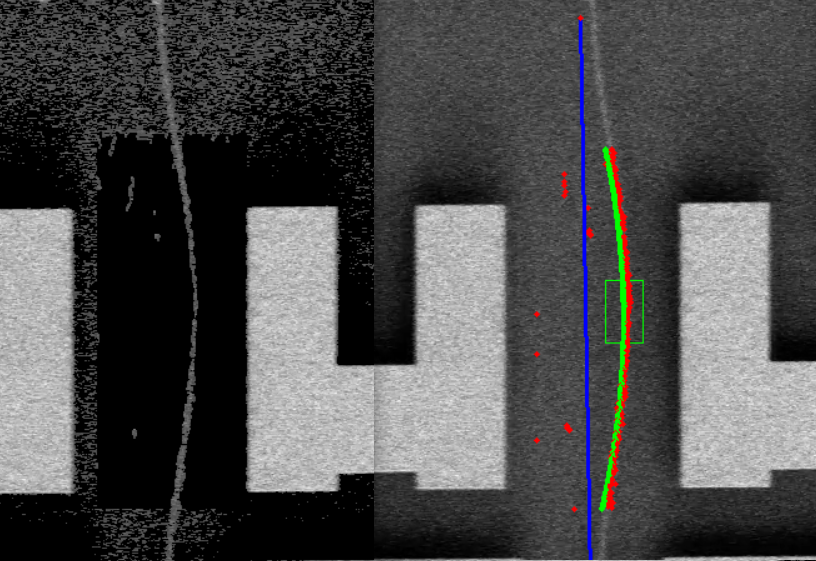}
    \end{center}
    \caption{Image denoising filter.}\label{fig:denoising}
\end{figure}

The beam appears blurred during its motion due to the low framerate (see Fig. \ref{fig:shadowing}). Directly applying the row-based maxima search method causes a large number of erroneous data point selections. Filtering some of the noise increases the performance of the algorithm by reducing the number of faulty points detected at the first stage of the algorithm (see Fig. \ref{fig:denoising}). This is done by setting the values of the pixels that do not have neighbouring bright pixels to zero. The underlying logic is that much of the noise is random, which means the area surrounding the noise might not be bright although the pixel value at the noise is high. On the other hand, the desired signal (beam) is a continuous object. Therefore, the surrounding area of a pixel within the beam should also be relatively bright as shown in Fig. \ref{sfig:method1}.

\renewcommand{\thefigure}{B\arabic{figure}}
\setcounter{figure}{0}
\section{\label{sec:curvefit}Nonlinear Curve Fitting of the Beam With the Gauss Newton Method}

Each data point has a unique value on the y-axis and the position of a given point on the beam on the x-axis is treated as the dependent variable. This relation can be represented as

\begin{equation}\label{eq:model}
    x_i = f(y_i) + e_i.
\end{equation}

\noindent{where $f$ is the governing solution, $x_i$ the deflection of the i$^{th}$ data point and $e_i$ is the error.}

Buckling analysis shows that for a doubly clamped beam under buckling load, the governing solution for deflection is given by

\begin{equation}\label{eq:governing}
    f(y_i) = c_1sin(ky_i) + c_2cos(ky_i) + c_3
\end{equation}

\noindent{which results in the equation below when Taylor expanded.}

\begin{eqnarray}\label{eq:taylor}
    f(y_i)_{j+1}=&&f(y_i)_j + \frac{\partial f(y_i)_j}{\partial c_1}{\Delta}c_1 + \frac{\partial f(y_i)_j}{\partial c_2}{\Delta}c_2\nonumber\\&&+ \frac{\partial f(y_i)_j}{\partial c_3}{\Delta}c_3\nonumber\\
    =&&f(y_i)_j + {\Delta}c_1sin(ky_i) + {\Delta}c_2cos(ky_i)\nonumber\\&&+ {\Delta}c_3
\end{eqnarray}

It must be noted that the parameter $k$ is not used in the least squares procedure even though it technically can be. However, including $k$ as a free parameter results in worse performance for our case since the model overfits or does not converge quickly (or not at all in some cases). Therefore, the value of $k$ was chosen as $\frac{2\pi}{L}$ where $L$ is the length of the beam.

In matrix form, the Eq.~\eqref{eq:model} and ~\eqref{eq:taylor} can be represented as

\begin{equation}\label{eq:matrix1}
    \bm{D} = \bm{Z_j}\bm{{\Delta}C}+\bm{E}
\end{equation}

or more explicitly

\begin{equation}\label{eq:matrix2}
    {\begin{bmatrix}
        x_1-f(y_1)\\
        x_2-f(y_2)\\
        \vdots\\
        x_n-f(y_n)
        \end{bmatrix}} = {\begin{bmatrix}
        sin(ky_1) & cos(ky_1) & 1\\
        sin(ky_2) & cos(ky_2) & 1\\
        \vdots & \vdots & \vdots\\
        sin(ky_n) & cos(ky_n) & 1\\
        \end{bmatrix}} {\begin{bmatrix}
        {\Delta}c_1\\
        {\Delta}c_2\\
        {\Delta}c_3
        \end{bmatrix}} + \bm{E}
\end{equation}

\noindent{where the parameters are updated in each iteration by}
\begin{equation}\label{eq:update}
    c_{k,j+1}=c_{k,j}+{\Delta}c_k.
\end{equation}

The least squares solution of Eq.~\eqref{eq:matrix1} for $\bm{{\Delta}C}$ at each step is given by

\begin{equation}\label{eq:leastsquares}
    \bm{{\Delta}C} = (\bm{Z_j}^{T}\bm{Z_j})^{-1}(\bm{Z_j}^{T}\bm{D}).
\end{equation}

The iteration is repeated until the solution converges or the number of iterations reaches a predetermined stopping criterion. This whole overall procedure is the \textbf{Gauss-Newton method}, and it can be applied to other models as well \cite{Chapra2014}.

For simpler models, linear regression can be used instead, which is much easier and directly provides the result without iterations. However, linearizing the model might lead to worse mean square error as compared to nonlinear regression.

% Right before starting your bib
\makeatletter
% Based on code found in aps4-2.rtx
\def\bibsection{%
  \par
  \baselineskip26\p@
  \bib@device{\linewidth}{82\p@}%
  \nobreak\@nobreaktrue
  \addvspace{19\p@}%
  \par
}
\makeatother

%\nocite{*} % prints all entries in a bibliography
%\bibliography{references}% Produces the bibliography via BibTeX.

\providecommand{\noopsort}[1]{}\providecommand{\singleletter}[1]{#1}%

\end{document}